\documentclass[twocolumn,superscriptaddress,letter]{revtex4}%
\usepackage{amssymb}
\usepackage{color}
\usepackage{graphicx}
\usepackage{dcolumn}
\usepackage{bm}
\usepackage[header,title,page,titletoc]{appendix}
\usepackage{amsmath}
\usepackage{amsfonts}%
\providecommand{\U}[1]{\protect \rule{.1in}{.1in}}

\begin{document}
\title{Realizing Universal Majorana Fermionic Quantum Computation }
\author{Ya-Jie Wu}
\author{Jing He}
\author{Su-Peng Kou}
\thanks{Corresponding author}
\email{spkou@bnu.edu.cn}
\affiliation{Department of Physics, Beijing Normal University, Beijing 100875, China}
\date{March, 2013}

\begin{abstract}
Majorana-fermionic quantum computation (MFQC) was proposed by Bravyi and
Kitaev (See Ref.\cite{Kitaev}), in which a fault-torrent (non-topological)
quantum computer built from Majorana fermions may be more efficient than that
built from distinguishable two-state systems. However, till now people don't
know how to realize a MFQC in a physical system. In this paper we proposed a
possible realization of MFQC. We find that the end of a line-defect of p-wave
superconductor or superfluid on a honeycomb lattice will trap a Majorana zero
mode, which will become the starting point of MFQC. Then we show how to
manipulate Majorana fermions to do universal MFQC, which possesses unique
possibilities for high-level local controllability, individual addressing, and
readout of the quantum states of individual constituent elements by using
timely cold-atom technology.

\end{abstract}
\maketitle

\textit{Introduction}.---Quantum computer makes use of the principles of
quantum physics to enhance the computational power beyond what is attainable
by a traditional computer. Various designs have been proposed to build a
quantum computer, such as manipulating electrons in a quantum dot or phonon in
ion traps, cavity quantum electrodynamics, nuclear spin by nuclear magnetic
resonance techniques. However, the realization of low decoherence condition in
these schemes is a great challenge to achieve a true quantum computer. Because
the noise and decoherence are presumably caused by local interactions, people
try to avoid them by encoding quantum information non-locally. This leads to
so called topologically protected qubit\cite{ki1,ioffe,kou1}. In particular,
the topological quantum computation is proposed based on the manipulation of
non-Abelian anyons\cite{ki2,sar,free,sar1,ge}. The possible candidates may be
the fractional quantum Hall state at filling factor $\nu=5/2$ in ultra
high-mobility samples\cite{eis,xia}, the two dimensional (2D) chiral
$p_{x}+\mathrm{i}p_{y}$ superconductors\cite{da}, and the
s-wave-superconductor-topological-insulator systems\cite{al,fu}. Yet, people
cannot realize universal quantum computation just by braiding Ising-type
non-Abelian anyons. In addition, among these approaches, accurate manipulation
of single anyon remains a major difficulty and new techniques are expected to
overcome this drawback.

In this paper we turn to an alternative type of fault-torrent quantum
computation - Majorana-fermionic quantum computation (MFQC).
Majorana-fermionic quantum computation was proposed by Bravyi and
Kitaev\cite{Kitaev}, in which the qubits are characterized by fermion's
parity. A quantum computer built from Majorana fermions may be more efficient
than that built from distinguishable two-state systems. Although MFQC is not
the topological quantum computation, the quantum information with Majorana
fermions also intrinsically immunes to decoherence. Thus it is possible to do
universal quantum computation by manipulating the two-Majorana-fermion
interaction ($\gamma_{1}\gamma_{2}$) and the four-Majorana-fermion interaction
($\gamma_{1}\gamma_{2}\gamma_{3}\gamma_{4}$). We find that the end of a
line-defect of p-wave superconductor (SC) or superfluid (SF) on a honeycomb
lattice will trap a Majorana zero mode, which will become the the starting
point of MFQC\cite{HE}. Then we show how to manipulate Majorana fermions to do
universal MFQC and also discuss the errors. In the end, we design an
experiment scheme for the realization of the p-wave SF in a Bose-Fermi mixture
on an optical honeycomb lattice.

\textit{P-wave {superconductor/superfluid} for spinless fermions on honeycomb
lattice}: We start from a spinless fermion model on a honeycomb lattice with
nearest attractive interaction as
\begin{equation}
\hat{H}=-t\sum_{\left \langle ij\right \rangle }\hat{c}_{i}^{\dag}\hat{c}%
_{j}-t^{\prime}\sum_{\left \langle \left \langle ij\right \rangle \right \rangle
}\hat{c}_{i}^{\dag}\hat{c}_{j}+h.c.-\mu \sum_{i}\hat{n}_{i}-U\sum_{\left \langle
ij\right \rangle }\hat{n}_{i}\hat{n}_{j}, \label{Ham}%
\end{equation}
where $t$ ($t^{\prime}$) and $U$ denote the strength of nearest (next nearest)
neighbor hopping and the nearest neighbor attractive interaction respectively,
$\mu$ is chemical potential, and $\hat{n}_{i}=\hat{c}_{i}^{\dag}\hat{c}_{i}$
denotes the particle number operator.

\begin{figure}[ptb]
\includegraphics[width=0.5\textwidth]{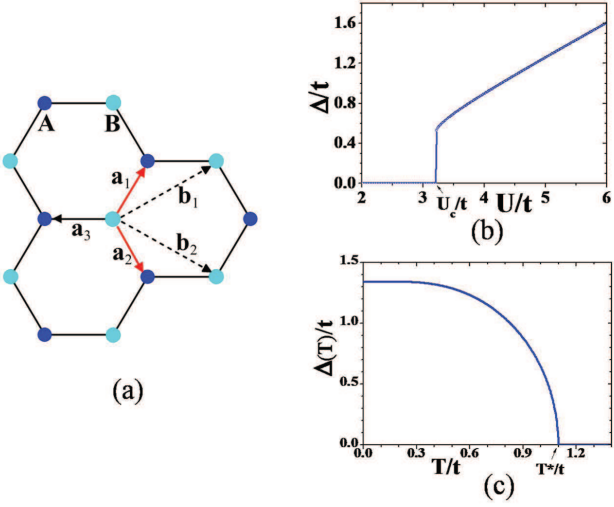}\caption{(a) The illustration
of honeycomb lattice of two interpenetrating triangular lattice. $A$ and $B$
denote $A$ and $B$ sublattice. The red links represent the paring bonds, on
which the pairing order parameter is finite; (b) The order parameter $\Delta$
vs. the interaction strength $U$ for the case of $t^{\prime}=0.01t$ at
half-filling; (c) The order parameter $\Delta$ vs. the temperature $T$ for the
case of $t^{\prime}=0.01t$, $U=5.24t$ at half-filling.}%
\label{hon}%
\end{figure}

As increasing the interaction strength, the system turns into a p-wave SC/SF
ground state in the region of $3.36t\leq U\leq7.12t$ for the case of
$t^{\prime}=0$ at zero temperature\cite{POL}. In the limit of $t^{\prime}\ll
t$, the p-wave SF order is still stable. For the p-wave SC/SF, the p-wave
pairing order parameter is given as $\mathbf{\Delta}_{j,j+\mathbf{a}_{1}%
}=\left \langle \hat{c}_{j}^{\dag}\hat{c}_{j+\mathbf{a}_{1}}^{\dagger
}\right \rangle =\Delta,$ $\mathbf{\Delta}_{j,j+\mathbf{a}_{2}}=\left \langle
\hat{c}_{j}^{\dag}\hat{c}_{j+\mathbf{a}_{2}}^{\dagger}\right \rangle =-\Delta,$
$\mathbf{\Delta}_{j,j+\mathbf{a}_{3}}=\left \langle \hat{c}_{j}^{\dag}\hat
{c}_{j+\mathbf{a}_{3}}^{\dagger}\right \rangle =0,$ where $\mathbf{a}_{\alpha}$
denotes a vector that connects nearest neighbor $i$ site and $i+\mathbf{a}%
_{\alpha}$ site as $\mathbf{a}_{1}=\frac{a}{2}(1,\sqrt{3}),$ $\mathbf{a}%
_{2}=\frac{a}{2}(1,-\sqrt{3}),$ $\mathbf{a}_{3}=a\left(  -1,0\right)  ,$
$\mathbf{b}_{1}=\mathbf{a}_{1}-\mathbf{a}_{3}$, $\mathbf{b}_{1}=\mathbf{a}%
_{2}-\mathbf{a}_{3}.\ $The lattice constant is defined as $a\equiv1$ in this
paper. In Fig.\ref{hon}(a), along red links, the SC/SF order parameter is
finite; while along black links, the SC/SF order parameter is zero. The
effective Hamiltonian then becomes
\begin{align}
H_{\mathrm{eff}}  &  =-t\sum_{\left \langle ij\right \rangle }\hat{c}_{i}^{\dag
}\hat{c}_{j}-t^{\prime}\sum_{\left \langle \left \langle ij\right \rangle
\right \rangle }\hat{c}_{i}^{\dag}\hat{c}_{j}-\sum_{j\in B}\mathbf{\Delta
}_{j,j+\mathbf{a}_{1}}\hat{c}_{j+\mathbf{a}_{1}}^{\dag}\hat{c}_{j}^{\dag
}\nonumber \\
&  -\sum_{j\in B}\mathbf{\Delta}_{j,j+\mathbf{a}_{2}}\hat{c}_{j+\mathbf{a}%
_{2}}^{\dag}\hat{c}_{j}^{\dag}+h.c-\mu \sum_{i}\hat{c}_{i}^{\dag}\hat{c}%
_{i}-U\sum_{i}\hat{n}_{i}\hat{n}_{i+\mathbf{a}_{3}}. \label{sc}%
\end{align}

At zero temperature, there exists a first order phase transition between
semi-metal and the SC/SF phase, where $U_{c}\simeq3.23t$ when $t^{\prime
}=0.01t$ at half filling as shown in Fig.1(b). In weakly interaction region,
$U<U_{c}$, the ground state is a semi-metal. When the interaction strength is
larger than $U_{c}$, in the region of $3.23t\leq U\leq7.12t,$ the ground state
turns into a p-wave SC/SF state. In the following parts, we focus on the case
of $U=5.24t$, of which the SC/SF order parameter $\Delta \simeq1.34t$ at zero
temperature. As the temperature increases, the order parameter decreases until
it disappears at the critical temperature $T^{\ast}=1.10t$ that corresponds to
breaking the Cooper pairing. The order parameter $\Delta$ versus temperature
$T$ is plotted in Fig.\ref{hon}(c) for the case of $t^{\prime}=0.01t,$
$U=5.24t$ at half-filling.

\begin{figure}[ptb]
\includegraphics[width=0.48\textwidth]{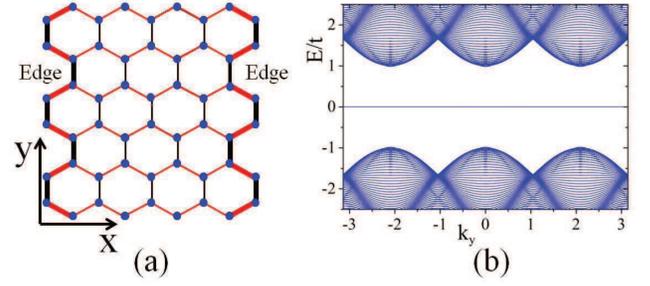}\caption{(Color online) Edge
states of p-wave SC/SF on honeycomb lattice for the system with armchair open
boundary. The parameters are chosen as $\Delta=1.34t$, $t^{\prime}=0.01t$,
$\mu=0.0136t$. Along red links, the SC/SF order parameter is finite; while
along black links, the SC/SF order parameter is zero.}%
\label{edge}%
\end{figure}

For the p-wave SC/SF on honeycomb lattice, there exists an energy gap for the
Bogliubov quasi-particles. For example, for the case of $U=5.24t$, the energy
gap $\Delta E_{g}$ is about $2.0t$. Remember that for the p-wave ($p_{x}$-wave
or $p_{y}$-wave) SC/SF state on a square lattice, the spectrum of
quasi-particles is always gapless. And it is obvious that this p-wave SC/SF on
honeycomb lattice is not a "topologically ordered" SC/SF. However, there exist
Majorana edge states. Now we consider the effective Hamiltonian in
Eq.(\ref{sc}) under periodic boundary condition along y-direction and opened
boundary condition (armchair boundary) along x-direction (the axis is defined
in Fig.\ref{edge}(a)).\textbf{ }Now\textbf{ }the p-wave SC/SF on a honeycomb
lattice behaves like coupled 1D p-wave SC/SFs along x-direction, where the
dangling Majorana fermions exist at the boundary\cite{Kitaev1}. Fig.\ref{edge}
shows the numerical results of the edge states for the case of $\Delta=1.34t$,
$t^{\prime}=0.01t$, $\mu=0.0136t$. From Fig.\ref{edge}, one can see that the
Majorana edge states always have a flat-band without hybridizing with each other.

\begin{figure}[ptb]
\includegraphics[width=0.5\textwidth]{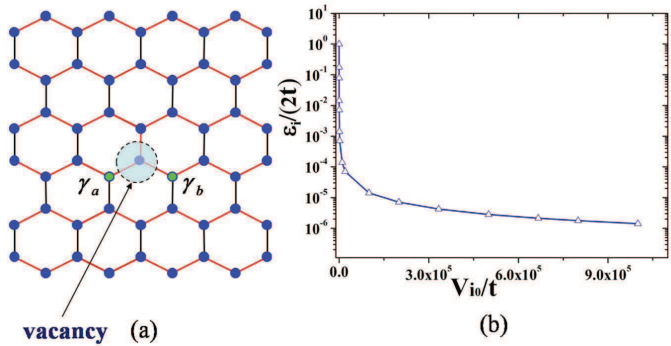}\caption{(a) The lattice
vacancy is represented by shaded-dashed-circle, and two green dots denote the
two Majorana modes $\gamma_{a},$ $\gamma_{b}$. Along red links, the SC/SF
order parameter is finite; while along black links, the SC/SF order parameter
is zero. (b) Half of the energy splitting $\varepsilon_{i}/t$ versus on-site
potential $V_{i_{0}}/t$. The detailed explicit Hamiltonian form is shown in
Eq.(\ref{sc}). Related parameters are chosen as $\Delta=1.34t,$ $t^{\prime
}=0,\mu=0$ at zero temperature. }%
\label{vac1}%
\end{figure}

\textit{Majorana zero modes around vacancies}: Now we study the properties of
a lattice defect for the p-wave SC/SF on honeycomb lattice in the presence of
an on-site potential at site $i_{0}$, $H_{\mathrm{eff}}\rightarrow
H_{\mathrm{eff}}+V_{i_{0}}\hat{c}_{i_{0}}^{\dagger}\hat{c}_{i_{0}}.$ After
solving the Bogolubov-de Gennes (BdG) equations with a single lattice defect
on a $50\times50$ lattice with periodic boundary condition, we obtain two
localized wave-function with energy $\pm \varepsilon$ around the lattice
defect. In the unitary limit, the lattice defect becomes a vacancy shown in
Fig.\ref{vac1}(a), of which we have an infinite on-site potential, i.e.,
$V_{i_{0}}\rightarrow \infty$. For the system with the particle-hole symmetry,
$\mu=0,$ $t^{\prime}=0$, two localized states turn into a pair of Majorana
modes with exact zero energy\cite{HE}, one Majorana mode at the left side of
the vacancy, and the other at the right side. See the results in
Fig.\ref{vac1}(a): there are two Majorana modes, $\gamma_{a}$, $\gamma_{b}$.
Away from the unitary limit, we have a finite on-site potential. As shown in
Fig.\ref{vac1}(b), the energy levels of the two localized modes versus
potential energy $V_{i_{0}}$ is plotted. From Fig.\ref{vac1}(b), we find that
as $V_{i_{0}}\rightarrow \infty,$ the energy of the localized state turns into zero.

\begin{figure}[ptb]
\includegraphics[width=0.45\textwidth]{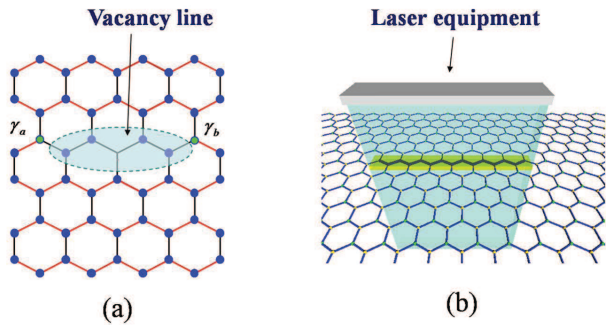}\caption{(a) The illustration
of a line-defect. Two Majorana modes $\gamma_{a},$ $\gamma_{b}$ are denoted by
green dots are localized at ends of the line-defect. (b) The illustration of
line-defect created by an addressing line-laser in optical lattice. Along red
links, the SC/SF order parameter is finite; while along black links, the SC/SF
order parameter is zero.}%
\label{ovacline1}%
\end{figure}

Next, we consider the case of a line of lattice-defect with multi-vacancy
along the SC/SF paring direction as shown in Fig.\ref{ovacline1}(a). Now the
Hamiltonian becomes $H_{\mathrm{eff}}\rightarrow H_{\mathrm{eff}}+\sum
_{k=1}^{N}V_{i_{k}}\hat{c}_{i_{k}}^{\dagger}\hat{c}_{i_{k}}.$ By numerical
calculations on a $50\times50$ lattice, from Fig.\ref{ovacline1}(a), we find
that there also exist two Majorana modes localized at both ends of the
line-defect, $\gamma_{a}$, $\gamma_{b}$. Now, the end of a line-defect can be
considered as the boundary of 1D p-wave SF. Thus each end of the line-defect
traps a dangling Majorana zero mode. When the line-defect is long enough, the
two Majorana modes decouple and have zero energy.

\textit{Universal Majorana Fermionic Quantum Computation}: Then we consider
the couplings between Majorana zero modes of vacancy-lines. For a general
case, the universal Hamiltonian of Majorana zero modes is given by
\begin{align}
H_{\mathrm{Majoaran}} &  =i\sum_{i}J_{i}\gamma_{ia}\gamma_{ib}+i\sum
_{\left \langle ij\right \rangle }J_{ij}^{(2)}\gamma_{ib}\gamma_{ja}\label{m}\\
&  +\sum_{\left \langle i,j\right \rangle }J_{ij}^{(4)}\gamma_{ia}\gamma
_{ja}\gamma_{ib}\gamma_{jb},\nonumber
\end{align}
where $J_{i}\ $denotes the coupling between two Majorana modes of a
line-defect $i$, $J_{ij}^{(2)}\ $denotes the coupling between two Majorana
modes of two different line-defects $i$, $j$, and $J_{ij}^{(4)}$ denotes the
strength of the four-Majorana particle interaction of two different
line-defects $i$, $j$. $\gamma_{ia}$ denotes a Majorana fermion at the left
side of a line-defect $i$ and $\gamma_{ib}$ denotes a Majorana fermion at the
right side of a line-defect $i$.

Therefore, the two Majorana modes $(\gamma_{ia},$ $\gamma_{ib})$ of a
line-defect form a physical fermion's parity qubit, $\left \vert 0\right \rangle
_{i}$ (quantum state with even fermion's parity) and $c_{i}^{\dag}\left \vert
0\right \rangle _{i}=\left \vert 1\right \rangle _{i}$ (quantum state with odd
fermion's parity), respectively. We label a pair of Majorana edge states
$(\gamma_{ia},$ $\gamma_{ib})$ by a complex fermion as $\gamma_{ia}=\left(
c_{i}+c_{i}^{\dag}\right)  ,$ $\gamma_{ib}=-i\left(  c_{i}-c_{i}^{\dag
}\right)  .$ By representing the Majorana fermions into complex fermions, the
universal Hamiltonian in Eq.(\ref{m}) of coupled Majorana zero modes becomes
\begin{align}
H_{\mathrm{Majoaran}} &  =\sum \limits_{i}J_{i}\left(  2c_{i}^{\dag}%
c_{i}-1\right)  \label{m2}\\
&  +\sum_{\left \langle ij\right \rangle }J_{ij}^{(2)}\left(  c_{i}c_{j}^{\dag
}-c_{i}^{\dag}c_{j}+c_{i}c_{j}-c_{i}^{\dag}c_{j}^{\dag}\right)  \nonumber \\
&  -\sum_{\left \langle i,j\right \rangle }J_{ij}^{(4)}\left(  2c_{i}^{\dag
}c_{i}-1\right)  \left(  2c_{j}^{\dag}c_{j}-1\right)  .\nonumber
\end{align}

We firstly calculate $J_{i},$ the coupling between two Majorana fermions at
the ends of\textit{ }a single lattice defect (a shortest line-defect with a
defect). In the unitary limit, we have two degenerate zero modes $(\gamma
_{ia},$ $\gamma_{ib})$ around the lattice defect for the SC/SF with
particle-hole symmetry ($\mu=0,$ $t^{\prime}=0$). Away from the unitary limit
of the SC/SF, $V_{0}\neq \infty,$ we have the finite energy splitting of the
Majorana modes as $\varepsilon_{i}\neq0.$ Then we can identify half of the
energy splitting $\varepsilon_{i}/2$ to be the coupling between two Majorana
modes around this lattice defect as $J_{i}\equiv \varepsilon_{i}/2$. When the
energy splitting $\varepsilon_{i}>0$ ($<0$), the system energetically favors
the empty (occupied) state. We also calculate the energy splitting of two
Majorana fermions of\textit{ }a line-defect with multi-defect (the number of
the defect is more than two) and found that the energy splitting of the two
Majorana modes is always zero. In this sense from Fig.\ref{vac1}(b), to tune
the coupling $J_{i}$ (that is $\varepsilon_{i}/2$) of a line-defect between
two Majorana modes $(\gamma_{ia},$ $\gamma_{ib})$, we can first shrink the
line-defect into a single vacancy and then change the on-site potential
$V_{0}$ from infinite to a finite value.

\begin{figure}[ptb]
\includegraphics[width=0.48\textwidth]{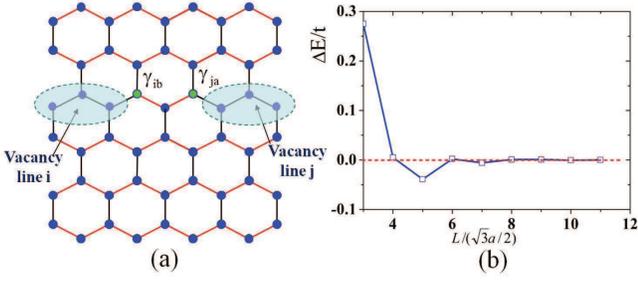}\caption{(a) The illustration
of two line-defects. The two Majorana fermions $\gamma_{ia},$ $\gamma_{ib}$
are denoted by the green dots that localize at ends of the line-defect 1 and
the line-defect 2, respectively. (b) The energy splitting $\Delta E/t$ versus
the distance between the ends of two line-defects as shown in (a). The
parameters are chosen as $\Delta=1.34t,$ $t^{\prime}=0.01t,$ $\mu=0.013t$.
Along red links, the SC/SF order parameter is finite; while along black links,
the SC/SF order parameter is zero.}%
\label{ovacline}%
\end{figure}

Next we calculate $J_{ij}^{(2)},$ the coupling between two Majorana modes at
the ends of two different line-defects. Now we consider two line-defects as
shown in Fig.5(a). When two line-defects are well separated, we have four
Majorana modes $\gamma_{ia},$ $\gamma_{ib},$ $\gamma_{ja},$ $\gamma_{jb}$ with
exact zero energy. Now we move the two line-defects nearby. The wave-functions
of the two Majorana zero modes $\gamma_{ib},$ $\gamma_{ja}$ will overlap and
give rise to a small energy splitting $\Delta E_{ij}$ from exact zero energy
level. The tunneling splitting $\Delta E_{ij}$ is about $J_{ij}^{(2)}\propto
e^{-\Delta E_{g}L_{i,j}}$ where $L_{i,j}$ is the distance between right-site
of $i$-th line-defect and left-site of the $j$-th line-defect. In
Fig.\ref{ovacline}(b), we plot the results of $J_{ij}^{(2)}$\ versus $L_{i,j}$
for the case of $t^{\prime}=0.01t$, $\mu=0.013t$. For this case $L_{i,j}>>a$,
we get zero tunneling amplitude as $J_{i,j}\rightarrow0.$ In this sense, to
tune the coupling $J_{ij}^{(2)}$ (that is $\Delta E_{ij}$) of two line-defects
between two Majorana modes $(\gamma_{ib},$ $\gamma_{ja})$, we can just change
the distance of two line-defects.

\begin{figure}[ptb]
\includegraphics[width=0.24\textwidth]{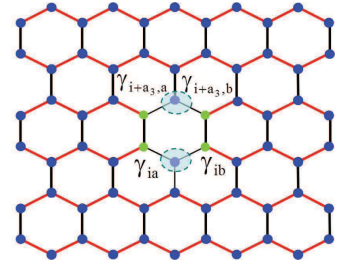}\caption{The illustration of
four Majorana modes induced by two defects that can combine into two complex
fermions ($c_{i},$ $c_{i+\mathbf{a}_{3}}$). Along red links, the SC/SF order
parameter is finite; while along black links, the SC/SF order parameter is
zero.}%
\label{ESP1}%
\end{figure}

Thirdly we calculate the strength of four-Majorana particle interaction
$J_{ij}^{(4)}$ between the four Majorana modes $\gamma_{ia},$ $\gamma_{ib},$
$\gamma_{ja},$ $\gamma_{jb}$ around two line-defects. Here we also consider
two line-defects in the unitary limit. When two line-defects are well
separated, there is no four-Majorana particle interaction at all. When two
line-defects are nearby, the four-Majorana particle interaction could be
induced by the nearest neighbor attractive interaction along un-pairing
direction as $-U\sum_{i}\hat{n}_{i}\hat{n}_{i+\mathbf{a}_{3}}$. As shown by
Fig.\ref{ESP1}, there are four Majorana fermions around the two line-defects
$\gamma_{ia},$ $\gamma_{ib},$ $\gamma_{i+\mathbf{a}_{3},a},$ $\gamma
_{i+\mathbf{a}_{3},b}$. The two zero modes $\gamma_{ia},$ $\gamma_{ib}$ form a
complex fermion $c_{i}$ by the definition $\gamma_{ia}=\left(  c_{i}%
+c_{i}^{\dag}\right)  ,$ $\gamma_{ib}=-i\left(  c_{i}-c_{i}^{\dag}\right)  $
and the other two zero modes $\gamma_{i+\mathbf{a}_{3},a},$ $\gamma
_{i+\mathbf{a}_{3},b}$ form another complex fermion $c_{j}$ by the definition
$\gamma_{i+\mathbf{a}_{3},a}=\left(  c_{i+\mathbf{a}_{3}}+c_{i+\mathbf{a}_{3}%
}^{\dag}\right)  ,$ $\gamma_{i+\mathbf{a}_{3},b}=-i\left(  c_{i+\mathbf{a}%
_{3}}-c_{i+\mathbf{a}_{3}}^{\dag}\right)  $. Then after considering the
nearest neighbor attractive interaction $-U\sum_{i}\hat{n}_{i}\hat
{n}_{i+\mathbf{a}_{3}},$ we will get residue interaction term between the two
complex fermions ($c_{i},$ $c_{i+\mathbf{a}_{3}}$). For this case, the
strength of the four-Majorana particle interaction term is given by
\begin{align}
-J_{i,i+\mathbf{a}_{3}}^{(4)} &  =\sum_{i}\psi_{i}^{\dagger}\psi
_{i+\mathbf{a}_{3}}^{\dagger}(H_{\mathrm{eff}})\psi_{i}\psi_{i+\mathbf{a}_{3}%
}\nonumber \\
&  =\sum_{i}\psi_{i}^{\dagger}\psi_{i+\mathbf{a}_{3}}^{\dagger}(-U\hat{n}%
_{i}\hat{n}_{i+\mathbf{a}_{3}})\psi_{i}\psi_{i+\mathbf{a}_{3}}\rightarrow
-\frac{U}{4}.
\end{align}
In this sense, to tune the four-Majorana particle interaction $J_{ij}^{(4)}$
of two line-defects between four Majorana modes $(\gamma_{ia},$ $\gamma_{ib},$
$\gamma_{ja},$ $\gamma_{jb})$, we can just move two parallel line-defects
nearby. However, for this case, the two-Majorana couplings between
($\gamma_{ia},$ $\gamma_{i+\mathbf{a}_{3},a}$) and ($\gamma_{ib},$
$\gamma_{i+\mathbf{a}_{3},b}$) are still finite, $J_{i,i+\mathbf{a}_{3}}%
^{(2)}\simeq t\neq0$.

By freely tuning all terms in Eq.(\ref{m2}), we may make the following two
universal gate set in fermionic quantum computation (up to phase factor) as
$e^{\frac{\pi}{8}\gamma_{i}\gamma_{j}}$ and $e^{i\frac{\pi}{4}\gamma_{1}%
\gamma_{2}\gamma_{3}\gamma_{4}}$\cite{Kitaev}: 1) To design $e^{\frac{\pi}%
{8}\gamma_{ia}\gamma_{ib}}$ gate we can shrink the line-defect into a single
vacancy and then fix the on-site potential $V_{0}$ under a given time-period
as $\Delta t=\frac{\pi \hbar}{8J_{ij}(V_{0})};$ 2) To design $e^{\frac{\pi}%
{8}\gamma_{ib}\gamma_{ja}}$ gate we can control the given two line-defects
nearby under fixed time-period as $\Delta t=\frac{\pi \hbar}{8J_{ij}%
^{(2)}(L_{i,j})};$ 3) To design $e^{i\frac{\pi}{4}\gamma_{ia}\gamma_{ib}%
\gamma_{i+\mathbf{a}_{3},a}\gamma_{i+\mathbf{a}_{3},b}}$ gate we can control
the given two line-defects nearby under fixed time-period as $\Delta
t=\frac{\pi \hbar}{4J_{i,i+\mathbf{a}_{3}}^{(4)}}$\cite{note}. Thus, in
principle, people are capable of doing arbitrary unitary transformation on the
protected subspace by controlling the vacancies.

\textit{Errors}: For the p-wave SF on honeycomb lattice, one always has a
large energy gap $\Delta E_{g}$ for the fermions and a very tiny energy
splitting $\Delta E$ of the nearly-degenerate Majorana modes, i.e., $\Delta
E\ll \Delta E_{g}$. Based on this condition ($\Delta E\ll \Delta E_{g}$), we may
ignore the excited states with $E>\Delta E_{g}$ and consider only the Majorana
modes as the protected sub-space. In this sense, we must consider the case at
fairly low temperature $k_{B}T\ll \Delta E_{g}$. Now there exist few excited
quasi-particles, $n_{f}\sim e^{-\Delta E_{g}/k_{B}T}$. In addition, the random
local potentials will lead to the energy splitting of the Majorana modes. Thus
when we move the vacancies nearby to do unitary operations, the random
energies of the Majorana modes will lead to errors. By numerical calculations,
we find that the energy splitting by random potential is very tiny. So we may
ignore this type of errors.

To store quantum information by the Majorana zero modes of the line-defect, we
must use a long line-defect to separate the Majorana modes. Thus, the distance
between two Majorana modes are set to be much bigger than lattice distance $a$
as $\min L_{i,j}>>a$. For this case, we get zero tunneling amplitude as
$J_{ij}^{(2)}(L_{i,j})\rightarrow0.$ Now the system truly immunes to decoherence.

\textit{Possible realization in cold atoms}: Now we propose an experiment
scheme for the realization of the spinless fermion model on a honeycomb
lattice with nearest attractive interaction given in Eq.(\ref{Ham}). To
realize this model, a Bose-Fermi mixture on optical honeycomb lattice of cold
atoms may be a possible candidate. As given in Ref.\cite{lew2}, the bosons
with unit filling form a Mott insulator and the spinless fermions with half
filling form a semi-metal. When the on-site repulsive interaction between
bosons and fermions $U_{bf}$ is larger than boson-boson repulsion $U_{bb}$
i.e., $U_{bf}>U_{bb},$ a spinless fermion and a bosonic hole bind into a
composite spinless fermion created by $\hat{c}^{\dagger}=\hat{f}^{\dagger}%
\hat{b}$, with $\hat{f}^{\dagger}$, $\hat{b}$ being the fermionic creation and
the\ bosonic annihilation operators, respectively. The effective nearest
attraction is given by $U/t=2\left(  U_{bf}/U_{bb}-1\right)  ,$ and the
effective nearest (next nearest) hopping $t=2t_{bf}^{2}/U_{bf}$ ($t^{\prime
}=2t_{bf}^{\prime2}/U_{bf}$), where $t_{bf}$ ($t_{bf}^{\prime}$) denotes the
tunneling strength for the bosons and fermions of the hopping between the
nearest (next nearest neighbor) neighbor lattices (here we have supposed that
$t_{b}=t_{f}=t_{bf}$, $t_{b}^{\prime}=t_{f}^{\prime}=t_{bf}^{\prime}$ for
simplicity). See detailed calculations in the supplementary materials. In
experiment, for example, $^{171}\mathrm{Yb-}^{174}\mathrm{Yb}$ or
$^{6}\mathrm{Li-}^{7}\mathrm{Li}$ mixture, the nearest neighbor effective
attracting interaction $U$ may be tuned employing well-characterized
homonuclear and heteronuclear Feshbach resonances\cite{fesh,Kit,Kem}. Thus in
cold atom system, people have ability to tune the parameters in Eq.(1)
including $t^{\prime}/t$, $U/t$.

On the other hand, recently, in cold atoms people have made great progress on
the manipulation of a single atom on an optical lattice\cite{chin,bloch1,weit}%
. A tightly focused dimple laser beam with an accuracy about $0.1a$ can be
applied onto individual lattice sites to generate a vacancy in the optical
lattice or line-vacancy\cite{chin,bloch1,weit}. By moving the addressing laser
beam as shown in Fig.4(b) from one site to another, people are able to
accurately manipulate the position and the length of a line-defect on a
two-dimensional optical lattice. Therefore two dimensional ultrocold atomic
Fermi gases provide an ideal platform for creating and manipulating Majorana fermions.

\textit{Conclusion}: We have proposed a feasible scheme towards designing a
quantum computer,{\ which incorporates intrinsic fault tolerance by
controlling the Majorana fermions accurately via controlling addressing laser
beams on optical lattice. }In particular, we find that the lattice defects
here plays the role of a Majorana fermion. Therefore, by creating and
manipulating lattice defects accurately via {addressing laser beams}, we can
do quantum computation fully optically. So we call this approach
"\emph{Majorana-fermionic quantum computation}" to emphasize the difference
between the topological quantum computation by braiding non-Abelian anyons.

\begin{acknowledgments}
This work is supported by NFSC Grant No. 11174035, National Basic Research
Program of China (973 Program) under the grant No. 2011CB921803, 2012CB921704.
\end{acknowledgments}


\begin{thebibliography}{99}
\bibitem {note}To eliminate the effects of the two-Majorana couplings between
($\gamma_{ia},$ $\gamma_{i+\mathbf{a}_{3},a}$) and ($\gamma_{ib},$
$\gamma_{i+\mathbf{a}_{3},b}$), after doing the $e^{i\frac{\pi}{4}\gamma
_{ia}\gamma_{ib}\gamma_{i+\mathbf{a}_{3},a}\gamma_{i+\mathbf{a}_{3},b}}$ gate,
we must do another two operations, $e^{\frac{7\pi}{4}\gamma_{ia}%
\gamma_{i+\mathbf{a}_{3},a}}$ gate and $e^{\frac{7\pi}{4}\gamma_{ib}%
\gamma_{i+\mathbf{a}_{3},b}}$ gate.
\end{thebibliography}
\end{document}